\documentclass[a4paper,fleqn,usenatbib]{mnras}
\usepackage[T1]{fontenc}
\usepackage{ae,aecompl}
\usepackage{graphicx}	
\usepackage{amssymb}
\usepackage{amsmath}	
 \usepackage{times}

\title[Empirical constraints on galaxy evolution]{New empirical constraints on 
the cosmological evolution of gas and stars in galaxies}

\author[Padmanabhan \& Loeb]{{Hamsa Padmanabhan$^{1}$\thanks{Email: 
hamsa@cita.utoronto.ca}
    and Abraham Loeb$^{2}$\thanks{Email: aloeb@cfa.harvard.edu}} \\
  $^{1}$ Canadian Institute for Theoretical Astrophysics,
60 St. George Street, Toronto, ON M5S 3H8, Canada\\
  $^{2}$ Astronomy department, Harvard University,
60 Garden Street, Cambridge, MA 02138, USA}

\date{Accepted ---. Received ---; in original form ---}

\pubyear{2020}

\begin{document}
\label{firstpage}
\pagerange{\pageref{firstpage}--\pageref{lastpage}}
\maketitle

\begin{abstract}
We combine the latest observationally motivated constraints on stellar 
properties in dark matter haloes, along with data-driven predictions for the 
atomic (HI) and molecular (H$_2$) gas evolution in galaxies, to derive empirical 
relationships between the build-up of galactic components and their evolution over 
cosmic time. {At high redshift ($z \gtrsim 4$), the frameworks imply that galaxies acquire their cold gas (both atomic and molecular) mostly by accretion, with the fraction of 
cold gas reaching about 20\% of the cosmic baryon 
fraction.} We {infer} a strong dependence of the star formation rate 
on the H$_2$ mass, suggesting a near-universal depletion timescale of 
0.1-1 Gyr in Milky Way sized haloes (of masses $10^{12} \ M_{\odot}$ at $z = 
0$). {There is also} evidence for a near-universality of the Kennicutt-Schmidt 
relation across redshifts, with very little dependence on stellar mass, if a 
constant conversion factor ($\alpha_{\rm 
CO}$) of CO luminosity to molecular gas mass  is assumed. Combining the atomic and molecular gas observations with the 
stellar build-up illustrates that galactic mass assembly in Milky-Way sized 
haloes proceeds from smooth accretion at high redshifts, towards becoming
merger-dominated at late times ($z \lesssim 0.6$). Our results can be used to constrain numerical
simulations of the dominant growth and accretion processes of galaxies over cosmic history.
\end{abstract}


\begingroup
\let\clearpage\relax
\endgroup

\section{Introduction}
The so-called `baryon cycle' in galaxies offers  novel insights 
into the inter-relationship between gas and stellar evolution across cosmic 
time. While we do not yet have a complete picture of the details of galaxy formation (for a review, see, e.g., \citet{naab2017}), some of the outstanding questions include: (i) the 
relative contributions of mergers and smooth accretion to the gas
assembly in galaxies \citep[e.g.,][]{keres2005, dekel2009, nelson2013}, as a function of halo mass and cosmic time, (ii) whether 
the physical processes governing star formation at high redshifts differ from 
those in the local universe, and (iii) the precise roles of atomic (HI) and 
molecular (H$_2$) gas \citep[e.g.,][]{daddi2010a, tacconi2013, saintonge2017} in driving the cosmological star formation rate.

Various semi-analytical and simulation methods have been used to predict the 
cosmological evolution of gas and stars in galaxies. While Semi-Analytical 
Methods \citep[SAMs; e.g.][]{guo2010, benson2012, somerville2008, popping2014, somerville2015} use 
sophisticated prescriptions having free parameters to model the physical 
parameters associated with the gas, stellar, black hole and radiation associated 
with galaxies, detailed hydrodynamical codes 
\citep[e.g.,][]{naab2017,tacchella2019} explicitly simulate these processes in a 
cosmological setting. Analytical techniques, such as toy models 
\citep[e.g.,][]{bouche2010, dekel2013, lilly2013} offer a complementary sketch for the 
stellar and gas build-up.

An alternative approach to constrain galaxy evolution is the empirical 
(data-driven) framework, in which observationally measured quantities are used 
to set direct constraints on the properties of gas and stars in haloes. For the
$\Lambda$CDM cosmological scenario, several empirical studies have placed 
constraints on the stellar properties in galaxies across cosmic time 
\citep[e.g.,][]{behroozi2013, moster2013, girelli2020, tacchella2018, rodriguez2017, behroozi2019} such as the stellar 
mass - halo mass (SHM) relation, using the technique of \textit{abundance 
matching}. Such methods have been extended to atomic and molecular gas in \citet{popping2015} by using physical prescriptions connecting gas profiles and stellar masses of galaxies. For atomic gas (HI) in galaxies, \citet{hparaa2017} used the 
combination of presently available data from galaxy surveys, HI intensity 
mapping experiments and Damped Lyman Alpha (DLA) observations to constrain an 
empirical HI mass - halo mass relation (HIHM). The inferred HIHM was found 
to be characterized by three free parameters and does not evolve strongly with 
redshift. In \citet{hpgk2017}, an 
equivalent, local HIHM was derived by matching the abundances of HI galaxies (at 
$z \sim 0$) observed in the HIPASS survey \citep{zwaan05} to the mass function 
of dark matter haloes \citep{sheth2002}. \footnote{Combining the HIHM so derived 
with the SHM obtained by \citet{moster2013} led to an inferred HI-stellar 
evolution which was consistent with various $z = 0$ measurements: from the Galex 
Arecibo SDSS Survey \citep[GASS;][]{catinella2010, catinella2013}, COLD GASS 
\citep{saintonge2011a, saintonge2011, catinella2012}, the HERA CO Line 
Extragalactic Survey  \citep[HERACLES;][]{leroy2009} and The HI
Nearby Galaxy Survey \citep[THINGS;][]{walter2008}.} 

The primary observational tracer of molecular gas (H$_2$) is carbon monoxide 
(CO) which is strongly connected to the star formation rate. In \citet{hpco}, 
constraints on the CO luminosity function at low redshifts \citep{keres2003} were
combined with intensity mapping observations at $z\sim 3$ from the CO Power 
Spectrum Survey \citep[COPSS;][]{keating2016} to predict the evolution of the CO 
luminosity - halo mass ($L_{\rm CO} - M$) relation via abundance matching. The 
observations were found to be consistent with a well-defined $L_{\rm CO} - M$ 
having four free parameters, motivated by the empirical SHM of 
\citet{moster2013}. In contrast to the HIHM, the inferred $L_{\rm CO} - M$ is 
observed to show a significant evolution across $z \sim 0-4$. 

The extended Press-Schecter (EPS) formalism provides a convenient framework to 
compute the assembly history of a given dark matter halo via mergers of smaller  haloes \citep{sheth2002}. Merger trees computed from numerical simulations 
\citep[e.g.,][]{klypin2011} allow the tracking of the most massive (main) 
progenitor halo for a given halo across cosmic time. In this paper, we combine 
the empirically determined prescriptions connecting the stellar \citep{behroozi2013, behroozi2019}, atomic \citep{hparaa2017} and 
molecular \citep{hpco} gas in galaxies to haloes, with the merger tree framework that 
describes halo assembly, to provide an understanding of the growth histories of 
the various galactic components and their dependences on each other. This 
analysis extends previous work to construct `baryon progenitor trees' which are 
directly motivated by observations. In so doing, it sheds light into the 
relative contributions of mergers and smooth accretion to gas mass assembly, and 
the dependence of the star formation history on the atomic and molecular gas 
depletion timescales. \footnote{Assembly bias and environmental effects are 
expected to be have a small to negligible effect on the halo models for baryonic 
gas, whose spatial extents are much smaller than the dark matter virial 
radius.}
Being completely empirical, this study is free from the uncertainties involved 
in physical models of stellar and gas evolution in galaxies. As such, it provides an important 
benchmark for calibrating the detailed physics in current and forthcoming 
simulations of galaxy formation, and enables the understanding of the dominant 
processes involved therein. 

The outline of this paper is as follows. In Sec. \ref{sec:halomodel}, we present the formalism which associates gas to dark matter haloes, which we connect to the empirical framework for the evolution of the stellar component in Sec. \ref{sec:buildup}. Finally, Sec. \ref{sec:results} summarizes our main results. 

\section{Halo model frameworks for atomic and molecular gas}
\label{sec:halomodel}

In this section, we briefly review the existing empirical frameworks developed 
for associating atomic and molecular gas to dark matter haloes.
For atomic gas (HI), we use the halo model for cosmological neutral hydrogen  
\citep{hparaa2017}, which combines constraints from HI galaxy surveys at $z \sim 
0$ \citep{zwaan05, zwaan2005a, martin10, martin12, braun2012}, intensity mapping 
experiments \citep[around $z \sim 1$; e.g.][]{switzer13} and the statistics of 
Damped Lyman Alpha (DLA) systems \citep[column density distributions, incidences 
and three dimensional clustering:][]{rao06,prochaska2009, 
noterdaeme12,fontribera2012, zafar2013} across $z \sim 0-5$. The results of a 
joint fit to all these datasets favour a well-defined mean HI mass - halo mass 
relation:
\begin{eqnarray}
M_{\rm HI} (M,z) &=& \alpha_{\rm HI} f_{\rm H,c} M \left(\frac{M}{10^{11} 
h^{-1} 
M_{\odot}}\right)^{\beta} \nonumber \\
&\times& \exp\left[-\left(\frac{v_{c0}}{v_c(M,z)}\right)^3\right] \, ,
\end{eqnarray}
with the three parameters (i) $\alpha_{\rm HI} = 0.09 \pm 0.01$, which denotes 
the 
average HI fraction relative to cosmic $f_{\rm H,c}$, (ii)   $\beta = -0.58 \pm 
0.06$, the logarithmic slope of the relation which represents the deviation from 
linearity of the prescription, and (iii) $v_{c0}$, given by $\log (v_{c,0}/ {\rm km \ s^{-1}}) = 1.58 
\pm 0.04$ which denotes the minimum virial 
velocity below which haloes preferentially do not host HI. {The halo mass function used for this purpose is that of \citet{sheth2002}.}

To describe the molecular gas (H$_2$) evolution, we use the results of 
\citet{hpco}, which infers a CO luminosity - halo mass relation having the 
physically motivated form:
\begin{equation}
L_{\rm CO} (M, z) = 2N(z) M [(M/M_1(z))^{-b(z)} + (M/M_1(z))^{y(z)}]^{-1} \, ,
\label{mosterco}
\end{equation}
with the parameters $M_1(z)$, $N(z)$, $b(z)$ and $y(z)$ themselves consisting of 
two terms - a constant term which describes the behaviour at $z \sim 0$, and an 
evolutionary component:
\begin{eqnarray}
\log M_1(z) &=& \log M_{10} + M_{11}z/(z + 1); \nonumber \\
N(z) &=& N_{10} + N_{11}z/(z + 1); \nonumber \\
b(z) &=& b_{10} + b_{11}z/(z + 1); \nonumber \\
y(z) &=& y_{10} +  y_{11}z/(z + 1) \, .
\label{comoster}
\end{eqnarray}
The best fitting values for these parameters, found from fitting the 
observations of \citet{keres2003} at $z \sim 0$ are given by $M_{10} = (4.17 \pm 
2.03) \times 10^{12} M_{\odot}, N_{10} = 0.0033 \pm 0.0016 \ \mathrm{K \ km/s  \ 
pc}^2 M_{\odot}^{-1}, b_{10} = 0.95 \pm 0.46, y_{10} = 0.66 \pm 0.32$.  The 
evolutionary components, derived by subsequently matching the COPSS results 
\citep{keating2016} at $ z \sim 3$ are given by:
\begin{eqnarray}
M_{11} &=& -1.17 \pm 0.85 ;                                                        
 \nonumber \\
N_{11} &=& 0.04 \pm 0.03 ;                                                                                                           
 \nonumber \\
b_{11} &=& 0.48 \pm 0.35 ; \nonumber \\
 y_{11} &=& -0.33 \pm 0.24 \, .                                                                                                                     
 \end{eqnarray}
The above framework can be converted into an equivalent H$_2$ mass to halo mass 
evolution by assuming a CO luminosity - H$_2$ mass conversion factor. 
The value of this factor -- denoted by $\alpha_{\rm CO}$, and defined through 
$M_{{\rm H}_2}  = \alpha_{\rm CO} L_{\rm CO}$ (with $M_{{\rm H}_2}$ in units of $M_{\odot}$ and $L_{\rm CO}$ in units of K km/s pc$^{-2}$) -- and its evolution are 
still observationally uncertain. Several studies \citep[e.g.,][]{bolatto2013} 
advocate the present value of $\alpha_{\rm CO}$ to be of order unity, and recent 
ALMA evidence  \citep[e.g.][]{cortese2017} indicating a higher fraction of molecular gas 
at high redshifts compared to atomic, may be consistent with a non-varying 
$\alpha_{\rm CO}$. Throughout this work, we assume $\alpha_{\rm CO} = 0.8$ 
across all redshifts under consideration.\footnote{A decreasing trend of 
$\alpha_{\rm CO}$ with $z$ may be advocated by the observational results of 
\citet{carleton2017} - which find about a $\sim 1$ dex decline in $\alpha_{\rm 
CO}$ between redshifts 0 and redshifts 3-4 \citep[see also][]{bolatto2013}. 
Since the detailed behaviour of $\alpha_{\rm CO}$ with redshift is essentially 
unconstrained, we prefer to stick with a non-varying value of $\alpha_{\rm CO}$ 
in the present work.}

\begin{figure}
    \centering
    \includegraphics[width = \columnwidth]{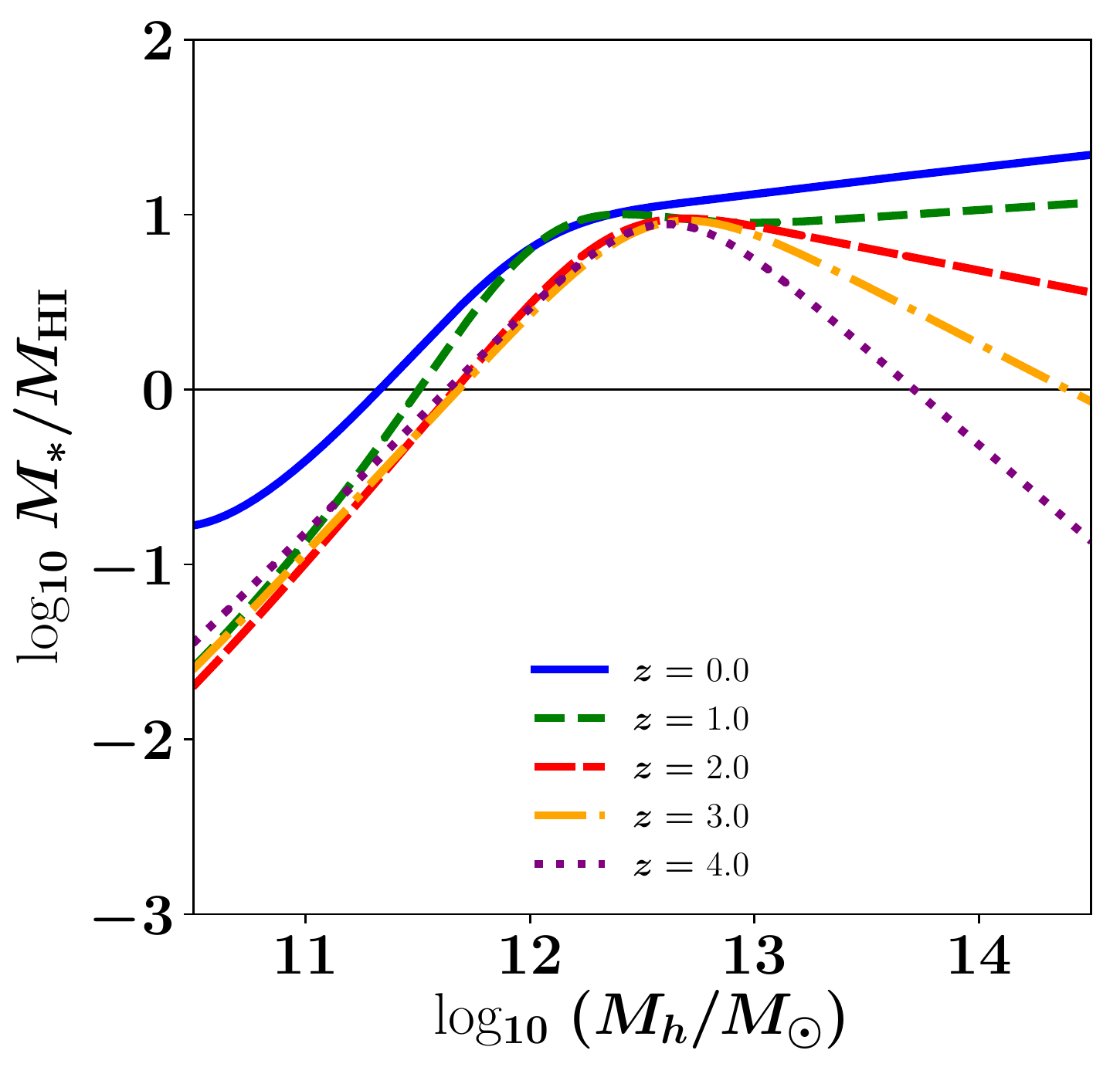}
    \caption{{Ratio of the average stellar mass $M_* (M_h, z)$ predicted by the SHM relation of \citet{behroozi2019} to the average HI mass, $M_{\rm HI}(M_h, z)$ predicted from the HIHM of \citet{hparaa2017} in host dark matter haloes as a function of their
halo masses $M_h$, for different redshifts $z = 0$ to 4. A ratio of unity (equal stellar and HI masses) is indicated by the black solid line.}}
    \label{fig:smhim}
\end{figure}

\section{Build-up of gas and stellar components}
\label{sec:buildup}

\subsection{Evolution of dark matter and stellar components}
\label{sec:dmstellar}

We can now combine the gas halo models outlined in Sec. \ref{sec:halomodel} with the existing results 
linking stellar properties in galaxies with their host dark matter haloes. For dark matter, we use the halo masses and  accretion histories from the compilation in \citet{behroozi2013}, which are based on the \textit{Bolshoi} simulations of 
\citet{klypin2011} with haloes identified using the {\textsc{ROCKSTAR}} halo finder 
\citep{behroozi2013b} and the corresponding merger trees \citep{behroozi2013c}. The \textit{Bolshoi} results were supplemented by those from two larger simulations: \citet[\textit{MultiDark};][]{riebe2011} and \citet[Consuelo;][]{behroozi2013c} for accuracy and resolution purposes. The combined simulation dataset allows to construct the trajectory of the most massive progenitor halo   
at each redshift for a given descendant halo at $z = 0$.

For the stellar component, we use the results of \citet{behroozi2019}, who combine observational data collected from the Sloan Digital Sky Survey (SDSS), the PRIsm 
MUlti-object Survey (PRIMUS), UltraVISTA, the Cosmic Assembly Near-infrared Deep 
Extragalactic Legacy Survey (CANDELS), and the FourStar Galaxy Evolution Survey 
(ZFOURGE) over  $0 < z < 10.5$ to derive empirical constraints on the stellar mass to halo mass relation across cosmic time. We also use the publicly available catalogs of empirically determined star formation histories from \citet{behroozi2013} (derived from a wide range of overlapping surveys over $z = 0$ to 8) for the evolution of the star formation rate across redshifts. 

\subsection{Connecting gas and stellar evolution}

We can now use the results of Sec. \ref{sec:halomodel} with the empirically derived stellar-halo mass relations to study the gas-galaxy connection across cosmic time. We begin by combining the empirical HIHM of \citet{hparaa2017} with the corresponding SHM derived by \citet{behroozi2019} to plot the ratio of the stellar mass to the HI mass ($M_*/M_{\rm HI}$) in galaxies, as a function of the host halo mass $M_h$ and redshift $z$, in Fig. \ref{fig:smhim}. The figure indicates that the ratio of HI to 
stellar mass is fairly independent of redshift and only depends on halo mass, {up to halo masses of $M_h \sim 10^{13} M_{\odot}$}.

It was found in 
\citet{hpgk2017} that the local ($z \sim 0$) HI-mass 
to stellar-mass ratio is about 25\% in
the rather broad range of halo masses from $10^{11}$ to $10^{13} M_{\odot}$ and 
decreases to about $10 \%$ at halo masses above this range. The differences between the present results at $z = 0$ and those of \citet[][see Fig. 2 of that paper]{hpgk2017}  in the low halo mass regime stem from the slightly different methodologies and datasets used to calibrate the HI-halo mass (HIHM) and stellar mass-halo mass (SHM) relations in \citet{hpgk2017}, relative to those used here. Specifically, it is known that at $M_{\rm h} \lesssim 10^{11} M_{\odot}$, abundance matching of HI gas in haloes from the HIPASS and ALFALFA datasets \citep{hpgk2017} predicts a somewhat lower (by $\sim 1$ dex) average HI mass fraction than a forward modelling MCMC-based approach to the HI observations at $z \sim 0-5$ \citep{hparaa2017}. On the other hand, the analysis of \citet{moster2013} used in \citet{hpgk2017} predicts a somewhat larger stellar mass ratio relative to that in \citet{behroozi2019} used here. The above two trends, taken together, lead to the observed differences between in the low halo mass regime, which are nevertheless within the scatter involved in the relevant SHM and HIHM relations (of the order of 10\% to 20\%). The SHM relation developed by other approaches, e.g. that of \citet{rodriguez2017} is also consistent with \citet{behroozi2019} at all halo masses in the range considered here (as illustrated in Fig. 34 of \citet{behroozi2019}), hence, using other SHM forms in the literature are also not expected to change the results significantly within the expected uncertainties. 

\subsection{Baryonic build-up: accretion and mergers}

\begin{figure}
 \begin{center}
 \includegraphics[width = 0.9\columnwidth]{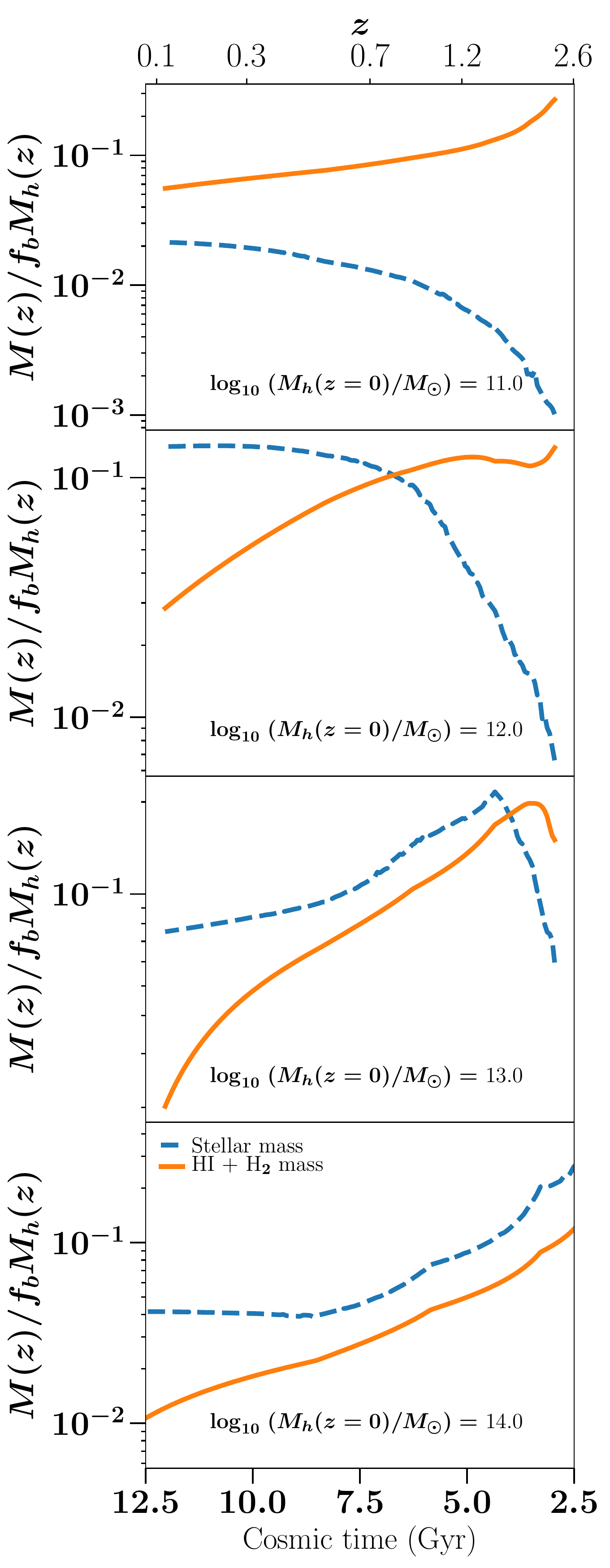} 
 \end{center}
\caption{{Evolution of the ratio of baryonic gas {(HI + H$_2$; orange solid lines) mass and stellar (blue dashed lines)  mass} to the progenitor host halo mass, relative to the cosmological baryon fraction ($f_b = \Omega_b/\Omega_m$). The plots assume a constant value of $\alpha_{\rm CO} = 0.8$ for 
converting CO luminosity to H$_2$ mass. From top to bottom, the panels show the evolution of these components with redshift $z$ in the most massive progenitors of dark matter haloes having masses $M_h (z = 0) = 10^{11}, 10^{12}, 10^{13}$ and $10^{14} M_{\odot}$ today.}}
\label{fig:HIH2masshalomass}
\end{figure}

\begin{figure}
 \begin{center}
\includegraphics[width = 
\columnwidth]{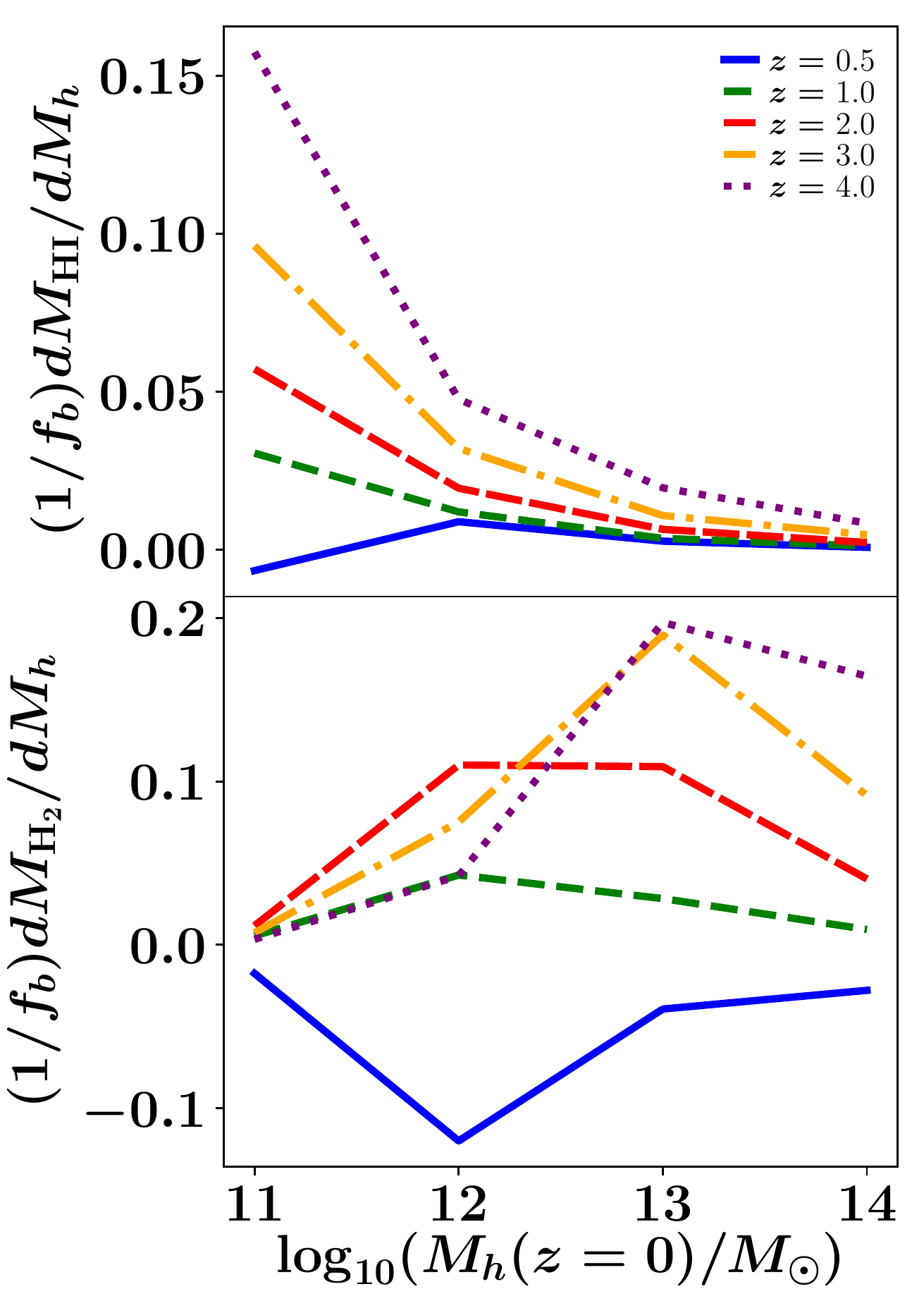}
 \end{center}
 \caption{{{Ratio of the differential change in cold gas mass to that in host halo mass}, $d M_{\rm HI}/ d M_h$ (top panel) and $d M_{{\rm H}_2}/ d M_h$ (lower panel) in the most massive progenitors of dark matter haloes, as a function of the descendent halo masses $M_h$ at $z = 0$. The ratios are indicated for redshifts $z = 0.5 - 4$ and normalized to the cosmic baryon fraction $f_b = \Omega_b/\Omega_m$. A constant value of $\alpha_{\rm CO} = 0.8$ is assumed for 
converting CO luminosity to H$_2$ mass.}}
 \label{fig:hih2buildup}
\end{figure}

\begin{figure}
 \begin{center}
  \includegraphics[width = \columnwidth]{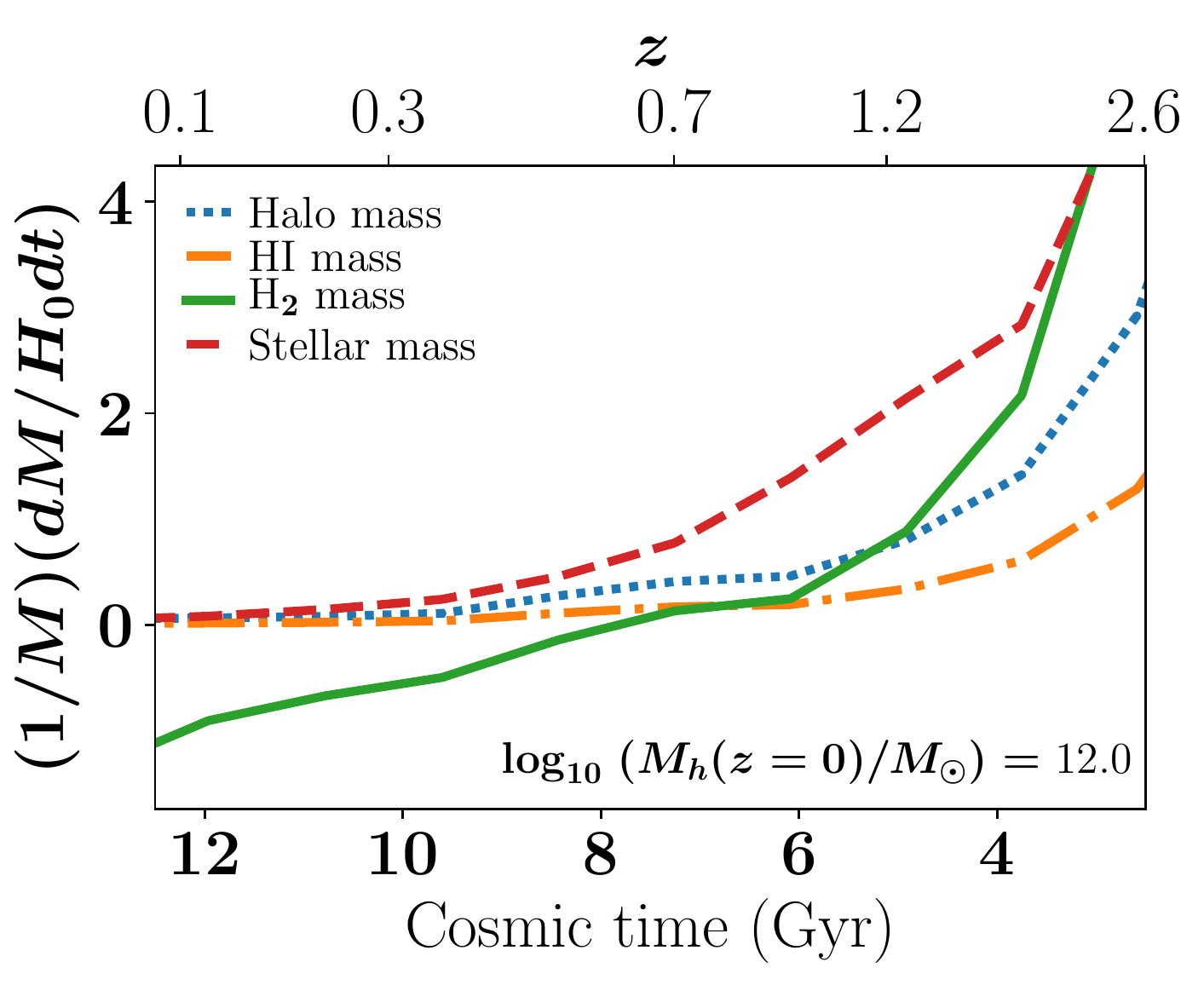} 
 \end{center}
 \caption{{Logarithmic derivatives of the progenitor halo ($d M_{\rm h}/M_{\rm h} H_0 dt$, blue dotted curve), HI ($d M_{\rm HI}/M_{\rm HI} H_0 dt$, orange dot-dashed curve),  H$_2$  ($d M_{\rm H_2}/M_{\rm H_2} H_0 dt$, green solid curve) and 
stellar ($d M_{*}/M_{*} H_0 dt$, red dashed curve) masses with respect to time, in a fiducial descendant dark matter halo of mass
$10^{12} M_{\odot}$ at $z = 0$.} The derivatives are normalized to the present-day Hubble constant. A constant ($\alpha_{\rm CO} = 0.8$) prescription is used to convert 
CO luminosity to molecular hydrogen mass.}
 \label{fig:accretion}
\end{figure}

Combining the dark matter merger trees described in Sec. \ref{sec:dmstellar} and the empirical SHM relation \citep{behroozi2019} leads to the evolution 
of the stellar mass trajectories across cosmic time. Analogous trajectories for the gas mass 
build-up can now be constructed by populating the dark matter merger trees with the gas-halo 
mass connections described in Sec. \ref{sec:halomodel}. Both these sets of trajectories
are shown  in Figs. \ref{fig:HIH2masshalomass} as a function of cosmic 
age, for four descendant dark matter halo masses ($10^{11} M_{\odot}$, $10^{12} 
M_{\odot}$, $10^{13} M_{\odot}$, and $10^{14} M_{\odot}$) at $z = 0$. Each figure shows the ratio of the stellar ($M_*(z)$, blue dashed lines) and gas   masses ($M_{\rm HI}(z) + M_{{\rm H}_2}(z)$, orange solid lines) to the progenitor halo mass $M_h(z)$, normalized to the cosmic baryon fraction $f_b = \Omega_b/\Omega_m$.  The plots assume a constant value of $\alpha_{\rm CO} = 0.8$ for 
converting CO luminosity to H$_2$ mass.

The 
figures show that the star formation efficiency is highest in haloes of masses $M_h (z = 0) \sim 10^{12} 
M_{\odot}$ today, with the stellar mass growing at the expense of the halo 
mass and reaching about 20\% of the cosmic baryon fraction by $z \sim 0$. At lower halo masses ($\sim 10^{11} M_{\odot}$), the gas depletion and stellar build-up show the same trend but the curves are shallower, {with the stellar and gas fractions reaching a few percent of the cosmic baryon fraction by $z \sim 0$}. At higher halo masses, the baryonic conversion efficiency is lower and decreases with cosmic time, and the proportion of neutral gas is smaller at $z \sim 0$ than for $10^{12} M_{\odot}$ haloes. {The stellar and gas fractions in haloes of masses $\sim 10^{13} - 10^{14} M_{\odot}$ today peaked around  $z \sim 2$, and account for between a few and 10 percent of $f_b$ by $z \sim 0$.} These results are broadly consistent with those found in earlier semi-empirical studies: \citet{behroozi2013} and \citet{popping2015}, who use different prescriptions to connect atomic and molecular gas to stellar masses in galaxies.

The two main modes of gas assembly in galaxies are by smooth accretion from the intergalactic medium (IGM) and 
mergers. Insight into smooth gas accretion may be gained by measuring what 
fraction of the baryon inflow turns into the atomic and molecular gas (HI and 
H$_2$) of the central galaxy. This is quantified in the panels of Fig. \ref{fig:hih2buildup}, which plot the ratios of the differential baryonic gas mass ($d M_{\rm HI}$, $d M_{{\rm H}_2}$) to the differential halo mass ($d M_h$), normalized to the cosmic baryon fraction $f_b$. The differential mass changes are calculated as the overall change in the gas and progenitor host halo masses taking place over an incremental redshift interval of $d \  \ln \ z = 0.01$.
The fractions $d M_{\rm HI}/d M_h$ and $d M_{\rm{H}_2}/d M_h$ are plotted for the most massive progenitor halo at each redshift with respect to the mass of the descendant halo at  $z = 0$. {The figures show 
that atomic gas accretion is dominant for $z \gtrsim 1$ at the lowest halo masses, and decreases with increasing halo mass. It is also seen (consistently with Fig. \ref{fig:HIH2masshalomass}) that  the consumption of molecular gas is most efficient for Milky-Way sized galaxies, as indicated by the dip in the H$_2$ accretion for halo masses $M_h \sim 10^{12} M_{\odot}$ at $z \sim 0$. For more massive galaxies, the molecular gas is largely accreted at high redshifts, reaching 
about 20 \% of the total baryon fraction. This is consistent with observations \citep[e.g.,][]{conselice2013} and theoretical predictions \citep[e.g.,][]{dekel2009} of accretion being the major driver of star formation in massive galaxies at $z \sim 1.5 - 3$.}

{Fig. \ref{fig:accretion} shows the logarithmic derivatives of the progenitor halo ($d M_{\rm h}/M_{\rm h} H_0 dt$, blue dotted curve), atomic ($d M_{\rm HI}/M_{\rm HI} H_0 dt$, orange dot-dashed curve), molecular ($d M_{\rm H_2}/M_{\rm H_2} H_0 dt$, green solid curve) and 
stellar ($d M_{*}/M_{*} H_0 dt$, red dashed curve) masses  with respect to time, in a descendant dark matter halo of mass
$10^{12} M_{\odot}$ at $z = 0$. The values are normalized to the present-day Hubble 
constant $H_0$. The figure serves to further illustrate the relative contribution of mergers and smooth accretion as a function of 
redshift.
The behaviour indicates that the stellar build-up follows the molecular gas at 
early times (where the green solid and red dashed curves are close to one another), but 
becomes merger-dominated at late times, following the halo mass evolution (where 
the red dashed curve approaches the blue dotted one)}. This is consistent with the 
simulations of \citet{lhuillier2012} and the observational evidences collected 
by \citet{sanchezalmeida2014}, both of which indicate that smooth accretion, 
rather than mergers, is the dominant growth mode for gas mass assembly 
in the majority of high-redshift galaxies ($z > 0.4$), whereas massive galaxies 
at lower redshifts are primarily merger-dominated. {The trends found in the present work are consistent with the average mass accretion rate into haloes found by \citet[][also discussed in \citet{lilly2013, 
dave2013}]{dekel2013} which are expected to hold for the baryonic 
accretion as well.}

\begin{figure}
 \begin{center}
  \includegraphics[width = \columnwidth]{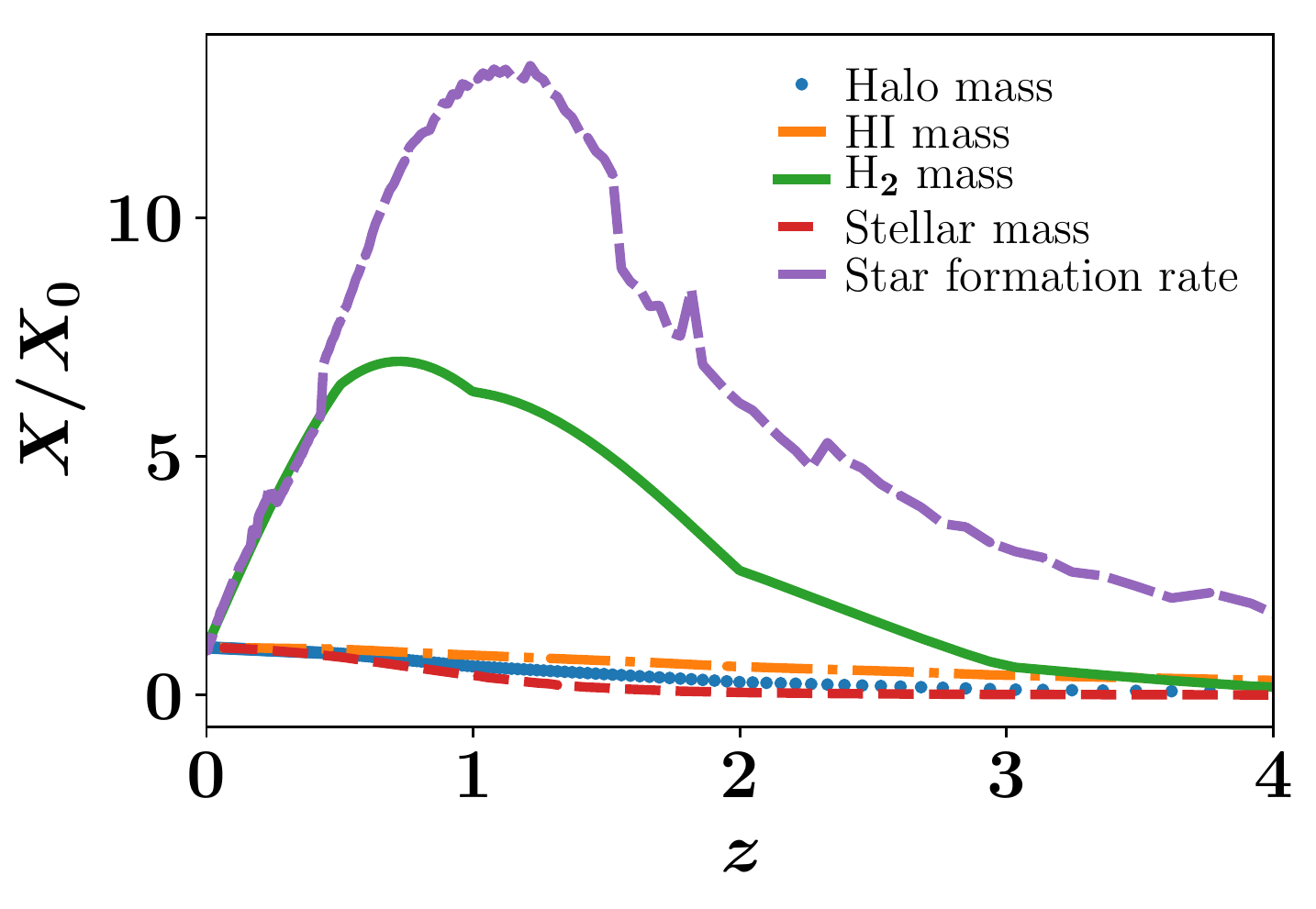}
 \end{center}
 \caption{{Ratios of the progenitor halo mass ($M_{\rm h}$, blue dotted curve), atomic HI mass ($M_{\rm HI}$, orange dot-dashed curve), molecular H$_2$ mass ($ M_{\rm H_2}$, green solid curve), 
stellar mass ($M_{*}$, red dashed curve) and star formation rate (purple long-dashed curve) at different redshifts to their present-day values, in a fiducial descendant dark matter halo of mass $M_h = 10^{12} M_{\odot}$ at $z = 0$.} A 
constant ($\alpha_{\rm CO} = 0.8$) prescription is used to convert CO luminosity to 
molecular hydrogen mass.}
 \label{fig:allgassfr}
\end{figure}

\subsection{Star formation and depletion timescale}

Finally, we explore {the connection between the empirically constrained star formation history \citep{behroozi2013}, to the build-up of the  various components (atomic, molecular, stellar 
and progenitor halo masses). }  {The ratios of each of these quantities: the progenitor halo mass ($M_{\rm h}$, blue dotted curve), HI mass ($M_{\rm HI}$, orange dot-dashed curve), H$_2$ mass ($ M_{\rm H_2}$, green solid curve), 
stellar mass ($M_{*}$, red dashed curve) and star formation rate (SFR, purple long-dashed curve) to their present-day values, 
are plotted in Fig. \ref{fig:allgassfr} as a function of redshift for a fiducial descendant halo of mass $10^{12} M_{\odot}$ at $z = 0$.} The figure
indicates that the H$_2$ mass (not the atomic HI mass) closely traces the star formation rate as a 
function of cosmic time, with the peak of the SFR build-up occurring slightly 
earlier than that of the H$_2$ fraction relative to the present. {This is consistent with observational findings that most star formation traces cold, dense gas and molecular clouds rather than warm atomic gas \citep[e.g.,][]{solomon1987, bolatto2008, bigiel2008, leroy2008}, as well as the predictions of the \textit{equilibrium growth/gas regulator models} developed in several theoretical studies \citep[e.g.,][]{bouche2010, lilly2013, dekel2014, peng2014}}.

\begin{figure}
    \centering
    \includegraphics[width = \columnwidth]{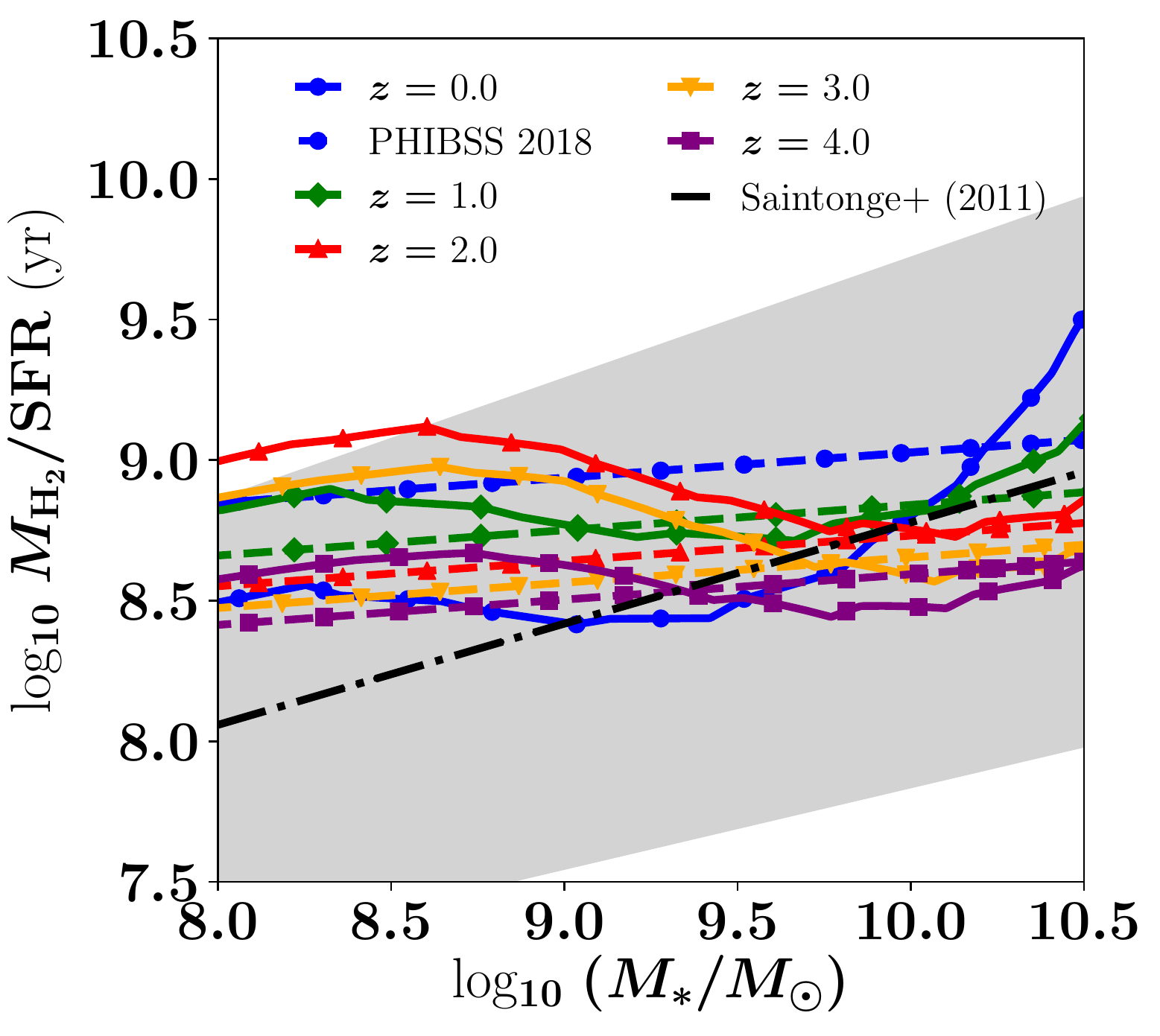}
    \caption{{Molecular gas depletion timescale, quantified by the ratio of H$_2$ mass to star formation rate, as a function of stellar mass at different redshifts. Solid linestyles show the results from the present work, and dashed linestyles show those obtained from the PHIBSS survey \citep{tacconi2018}. In both sets, the different colors and marker symbols correspond to the different redshifts: $z = 0$ (blue dots), $z = 1$ (green diamonds), $z = 2$ (red upward triangles), $z = 3$ (yellow downward triangles), and $z = 4$ (purple squares). The fitting function 
derived by \citet{saintonge2011a} at $z \sim 0$ (which is based on observations 
of $M_* > 10^{10} M_{\odot}$ galaxies) is plotted as the dot-dashed line, with its uncertainty range shown by the grey band.} A constant 
$\alpha_{\rm CO} = 0.8$ is assumed when converting CO luminosity to H$_2$ mass.}
    \label{fig:deptime}
\end{figure}

{The connection between star formation and molecular gas consumption can be well-quantified by the \textit{depletion time}, which is usually defined as the characteristic timescale for converting molecular gas into stars \citep[e.g.,][]{bigiel2008, Daddi2010, genzel2010, saintonge2011, tacconi2018, freundlich2019}. In Figure
\ref{fig:deptime}, we plot the mean depletion timescale (computed as $t_{\rm dep} = M_{\rm H_2}/{\rm SFR}$) 
 as a function of the stellar mass at different redshifts, compared to the results of observations. Solid color linestyles show the results from the present work, and dashed color linestyles show those obtained from the PHIBSS survey \citep{tacconi2018}. In both sets, the different colors and marker symbols correspond to the different redshifts: $z = 0$ (blue dots), $z = 1$ (green diamonds), $z = 2$ (red upward triangles), $z = 3$ (yellow downward triangles), and $z = 4$ (purple squares). The black dot-dashed line shows the fitting function 
derived by \citet{saintonge2011a} at $z \sim 0$, which is based on observations 
of $M_* > 10^{10} M_{\odot}$ galaxies, with its uncertainty range shown by the grey band.} 

 {The PHIBSS results find that the depletion timescale for galaxies along the star forming main sequence is well parametrized by a linear trend between $\log t_{\rm dep}$ and $\log M_*$, with  $t_{\rm dep} \propto \delta M_*^{0.09 \pm 0.05}$ where $\delta M_* = (M_*/5 \times 10^{10} M_{\odot})$. The observed weak dependence on the stellar mass is similar to the trend found in our present results. The PHIBSS observations also find a very shallow dependence with redshift, $t_{\rm dep} \propto (1+z)^{-0.62 \pm 0.13}$, indicating evidence for similar physical processes driving star formation at low and high redshifts. {The characteristic value of the depletion timescale found here, $t_{\rm dep} \sim 1$ Gyr, is also consistent with several other observational results \citep[e.g.,][]{bigiel2008, Daddi2010, genzel2010, tacconi2013}. and earlier semi-empirical work \citep{popping2015} that suggest an almost
universal depletion timescale across redshifts. At low galaxy masses, we find a somewhat different trend in the depletion timescale compared to the observations (which suggest a decreasing $t_{\rm dep}$ between $z \sim 0-2$). This could indicate evidence
 a lower value of $\alpha_{\rm CO}$ at higher redshifts compared to the
present \citep{bolatto2013}, and/or the impact of UV photodissociation in the low-metallicity ISM which can change the $\alpha_{\rm CO}$ at low galaxy masses \citep{leroy2013}.}
}

{
We also note that the presently adopted value of $\alpha_{\rm CO} = 0.8 M_{\odot}$/(K km/s pc$^2)$ is usually used for mergers, massive starbursts and particularly luminous galaxies above the main sequence. The observational results, on the other hand, use more complicated conversion factors which depend on metallicity (cf. Equation 2-4 of \citet{tacconi2018}, with the metallicity given by \citet{pettini2004}), and thus implicitly depend on stellar mass and redshift through the evolution of an assumed mass-metallicity relation (e.g., \citet{genzel2015}). Given the uncertainties in these relations and the existence of other factors influencing the resultant $\alpha_{\rm CO}$ (as also noted previously in footnote 3, Sec. \ref{sec:halomodel}), we have neglected any differences between the $\alpha_{\rm CO}$ used in the observations and that used in the present analysis (which does not account for changes in $\alpha_{\rm CO}$ with mass and redshift).
We note that using a Galactic $\alpha_{\rm CO}=4.36 M_{\odot}$/(K km/s pc$^2$) corrected for mass and metallicity, as advocated by the observations of \citet{tacconi2018, freundlich2019} would increase the tension with the observations, though this could have different possible causes. The conversion factor could indeed be overestimated, but the discrepancy could also be related to the representativity of the different samples used: those implied in the abundance matching relations, and those used by the \citet{saintonge2011, freundlich2019} and \citet{tacconi2018} studies -- which focus on galaxies on and around the main sequence, which are actively star forming and hence have much larger H$_2$ content. As such, the depletion times observed in these samples would be lower than in samples that also include quiescent galaxies. Given these considerations, we emphasize that Fig. \ref{fig:deptime} should be interpreted only as indicative of the average depletion timescales and their trends in evolution with mass and redshift. A detailed study of the variables influencing the CO-to-H$_2$ conversion factor across redshifts requires inputs from hydrodynamical simulations, and is left to future work.}

\section{Summary and discussion}
\label{sec:results}

We have developed several data-driven constraints on the build-up 
of atomic and molecular gas in galactic haloes, and their connections to 
observed stellar properties. This work brings together existing empirical 
constraints on the HI mass - halo mass relation and its evolution across 
redshifts \citep{hpgk2017, hparaa2017}, the luminosity and associated halo 
masses of CO-emitting galaxies across $z \sim 0-4$ \citep{hpco} to calculate 
molecular gas mass evolution, and the observationally motivated stellar - halo 
mass \citep{behroozi2019} and star formation rate across the same epochs calibrated by 
\citet{behroozi2013}. We find the following main results:

\begin{enumerate}
\item The mean stellar/HI mass ratio is almost universal with redshift. The 
dependence of this ratio with stellar mass is consistent with most observations
\citep[including, e.g., the latest findings from][at $z \sim 0$]{janowiecki2020}, and indicates that the underlying 
physics may be independent of redshift and only depend on halo mass. This 
points to mergers as a possible mode of stellar and atomic gas build-up, which 
is consistent with the predictions of theoretical models at low redshifts for 
massive galaxies \citep[e.g.,][]{dekel2009}.

\item At high redshifts, we find that most of the star formation is due to 
smooth accretion, rather than mergers, in Milky-Way sized haloes (of masses 
$10^{12} M_{\odot}$ at $z = 0$). This supports the `cold mode' of gas accretion 
at high redshifts predicted by theoretical models \citep[e.g.,][]{dekel2009}, and implies 
that most of the star formation is expected to take place in quiescent disk 
galaxies \citep[e.g.,][]{Daddi2010}, rather than merger-driven starbursts at these 
epochs. It is also, in turn, consistent with the picture of extended galactic 
disks based on rest-frame UV/optical \citep[e.g.,][]{bell2005, elbaz2007} and H$\alpha$ 
spectroscopy of galaxies \citep[e.g.,][]{genzel2008} and Damped-Lyman Alpha 
(DLA) system observations \citep[e.g.,][]{wolfe1986} at $z \gtrsim 2$. 

\item The star formation is strongly connected to the 
molecular gas (H$_2$) depletion timescale and negligibly to the atomic 
gas (HI). This reiterates the result, found in {several theoretical and observational studies \citep[e.g.,][]{solomon1987, bolatto2008, bigiel2008, leroy2008, bouche2010, lilly2013, dekel2014, peng2014}} as well as, e.g. the latest findings of \citet{wang2020} advocating 
the role of HI only as an `intermediary' in the process of star formation. It is also 
consistent with the arguments of \citet{prochaska2009} that point to a 
`self-correcting balance' in atomic gas: the HI replenishment from the 
intergalactic medium is compensated by its conversion to H$_2$ which is 
consumed by star formation. This, in turn, is linked to the observed constancy 
of  $\Omega_{\rm HI}$, the HI density parameter across redshifts as measured 
from DLA studies and 21 cm experiments (for a compilation of recent observations, see \citet{hptrcar2015} and references therein). {Closed box models for gas consumption \citep[e.g.,][]{bauermeister2010}, also advocate the intermediary role of HI in star formation, as coming from ionized gas in the IGM which is ultimately converted into H$_2$.}

 \item The depletion timescale for the consumption of molecular gas, quantified by $t_{\rm dep} 
= (M_{{\rm H}_2}/$SFR) is of the order of 0.1 - 1 Gyr, consistently with several 
observational results \citep[e.g.,][]{kennicutt1983, genzel2010, tacconi2018, freundlich2019} at $z\sim 0-2$. 
The $t_{\rm dep}$ does not depend strongly on stellar mass, which is also consistent 
with recent observations \citep[e.g.,][]{janowiecki2020,tacconi2018, freundlich2019} at low 
redshifts. The trend is predicted to hold at higher redshifts as well, 
suggesting a universality in the Kennicutt-Schmidt relation 
\citep{kennicutt1998}. Taken together with the observations of \citet{daddi2010a, 
tacconi2013} and \citet{genzel2010}, our findings may provide evidence for a decreasing 
CO-to-H$_2$ conversion factor at high redshifts as compared to its current
value. 

\end{enumerate}

{The key element in this work is the combination of abundance matching empirical relations for the atomic, molecular and stellar mass as a function of halo mass. However, these different components are not independent of each other. A useful way of quantifying the resulting uncertainty is to analyse the accuracy of the abundance matching hierarchy assumed for each component (stars, atomic gas, molecular gas) from the scatter in the respective relations. The typical uncertainties on the gas - halo relations are of the order of a few - 10\% at present (depending on redshift, the  scatter  on the halo model parameters are summarized in Table 3 of \citet{hparaa2017} and Table 1 of \citet{hpco}). For the stellar component, the typical scatter in the SHM may be about 10-20\% as illustrated in \citet{behroozi2010} for Milky-Way sized haloes. {The small scatter in the relations does not, however, imply they are independent; we also note that phenomena such as outflows resulting from feedback may invert the hierarchy assumed by the abundance matching technique (for example when depleting the gas).}}

The empirical constraints developed here serve as an important benchmark for 
calibrating the results of future simulations and semi-analytical models of 
galaxy formation that attempt to model the gas and stellar components in a 
self-consistent manner. {In future work, these techniques could be extended to the evolution of galaxy and HI disc sizes by combining the observations of stellar disks in, e.g., \citet{vanderwel2014} and the empirical evolution of the HI profile derived from \citet{hparaa2017} across redshifts, and exploring the consequences for star formation and ISM physics.}  Similar results can also be derived for various other dependent relations, including those exploring trends relative to the star-forming main sequence \citep[e.g.,][]{tacconi2018}. Forthcoming observations of atomic (e.g. with the SKA\footnote{https://www.skatelescope.org/}
and its precursors) and molecular gas (e.g. with the ALMA\footnote{https://almascience.nrao.edu/about-alma/alma-basics}/VLA\footnote{https://science.nrao.edu/facilities/vla/}), as well as 
gravitational lensing surveys detecting cosmic shear, will be useful to further 
constrain the physical processes involved to provide a complete picture of 
galaxy evolution.

\section*{Acknowledgements}
\addcontentsline{toc}{section}{Acknowledgements}
HP thanks Dongwoo Chung and Mubdi Rahman for useful clarifications related to 
the empirical studies of stellar properties in dark matter haloes. The work of AL was supported in part by the Black Hole Initiative at Harvard University, which is funded by grants from JTF and GBMF. {We thank the referee for a detailed and comprehensive report that significantly improved the content and quality of the presentation.}

\bibliographystyle{mnras}
\bibliography{mybib} 
\bsp	% typesetting comment
\label{lastpage}
\end{document}